\documentclass[aps,preprint,pre]{revtex4}
\usepackage{amsmath}
\begin{document}
\title{Nonlinear Higher-Order Thermo-Hydrodynamics: Generalized Approach in a
Nonequilibrium Ensemble Formalism}
\author{\'Aurea R. Vasconcellos} 
\affiliation{Instituto de F\'{\i}sica ''Gleb Wataghin'', Universidade
Estadual de Campinas - Unicamp, CP 6165, 13083-970, Campinas-SP, Brazil} 
\author{J. Galv\~ao Ramos}
\affiliation{Instituto de F\'{\i}sica ''Gleb Wataghin'', Universidade
Estadual de Campinas - Unicamp, CP 6165, 13083-970, Campinas-SP, Brazil} 
\author{Roberto Luzzi}
\affiliation{Instituto de F\'{\i}sica ''Gleb Wataghin'', Universidade
Estadual de Campinas - Unicamp, CP 6165, 13083-970, Campinas-SP, Brazil} 
\date{\today}
\begin{abstract}
Construction of a nonlinear higher-order thermo-hydrodynamics, including
correlations, in the framework of a Generalized Nonequilibrium
Statistical Grand-Canonical Ensemble is presented. In that way it is 
provided a particular formalism for the coupling and simultaneous 
treatment of the kinetics and hydrodynamic levels of description. 
It is based on a complete thermostatistical approach in terms of the 
densities of energy and of matter and their fluxes of all orders, as 
well as on their direct and cross correlations, covering systems 
arbitrarily driven-away-from equilibrium. The set of coupled nonlinear
integro-differential hydrodynamic equations is derived. Illustrations of 
the application of the theory are described in the follow up article.
\end{abstract}
\maketitle
\section{Introduction}
It has been noticed that one of the complicated problems of the
nonequilibrium theory of transport processes in dense gases and
liquids is the fact that their kinetics and hydrodynamics are
intimately coupled and must be treated simultaneously (E.g. see
Refs.[1-5]). Moreover, on this we may say that microscopic
descriptions of hydrodynamics, that is, associated to derivation of
the kinetic equations from classical or quantum mechanics and
containing kinetic (transport) coefficients written in terms of
correlation functions, is a traditional problem of long standing. 
An important aspect is the derivation of constitutive laws which
express thermodynamic fluxes (or currents as those of matter and
energy) in terms of appropriate thermodynamic forces (typically
gradients of densities as those of matter and energy). In their most
general form these laws are nonlocal in space and noninstantaneous in
time. The nonlocality is usually dealt with in terms of spatial
Fourier transforms, and then the laws - now expressed in reciprocal
space - become dependent on wave-vector ${\mathbf Q}$. A first
kinetic-hydrodynamic approach can be considered to be the so-called 
{\it classical (or Onsagerian) thermo-hydrodynamics}; it gives
foundations to, for example, the classical Fourier's and Fick's
diffusion laws. But it works under quite restrictive conditions,
namely, local equilibrium; linear relations between fluxes and
thermodynamic forces (meaning weak amplitudes in the motion) with
Onsager's symmetry laws holding; near homogeneous and static movement
(meaning that the motion can be well described with basically Fourier
components with long wavelengths and low frequencies, and then
involves only smooth variation in space and time); weak and rapidly
regressing fluctuations \cite{a6,a7}.

Hence, more advanced approaches require to lift these
restrictions. Consider first near homogeneity, which implies validity
in the limit of long wavelengths (or wavenumber $Q$ approaching zero),
and to go beyond it is necessary to introduce a proper dependence  on
$Q$ valid, in principle, for intermediate and short wavelengths
(intermediate to large wavenumbers). In phenomenological theories this
corresponds to go from classical irreversible thermodynamics to extended
irreversible thermodynamics \cite{a8,a9}. This is what has ben called
{\it generalized hydrodynamics}, a question extensively debated for
decades by the Statistical Mechanics community. Several approaches
have been used, and a description can be consulted in Chapter 6 of the
classical book on the subject by Boon and Yip \cite{a10}. Introduction
of nonlocal effects for describing motions with influence of ever
decreasing wavelengths, going towards the very short limit, has been
done in terms of expansions in increasing powers of the wavenumber,
which consists in what is nowadays sometimes referred to as 
{\it higher-order hydrodynamics} (HOH). Attempts to perform such 
expansions are the so-called Burnett and super-Burnett approaches in
the case of mass motion, and Guyer-Krumhansl approach  in the case of 
propagation of energy (see for example Refs.[11] and [12]). An usual 
approach has been based on the moments solution procedure of 
Boltzmann equation, as in the work of Hess \cite{a13}, using a 
higher-order Chapman-Enskog solution method. The Chapman-Enskog 
method provides a solution to Boltzmann equation consisting of a 
series in powers of the Knudsen number, $K_{n}$, given by the ratio 
between the mean-free path of the particles and the scale of change 
(relevant wavelengths in the motion) of the hydrodynamc 
fields. Retaining the term linear in $K_{n}$ there follows 
Navier-Stokes equation, the term in $K_{n}^{2}$ introduces Burnett-like 
contributions, and the higher-order ones ($K_{n}^{3}$ and up) the 
super-Burnett contributions. 

A satisfactory development of a HOH being also nonlinear and including
fluctuations is highly desirable for covering a large class of
hydrodynamic situations, and, besides its own scientific interest,
also for obtaining insights into technological-industrial processes
having an associated economic interest. Also we can mention its
fundamental relevance in Oceanography and Metereology, as for example,
the study of quite revelant phenomena as thermohaline circulation and
ENSO (El Ni\~no Southern Oscillation); see for example Refs.[14] and
[15] respectively. Moreover, it has been stated \cite{a16} that the
idea of promoting hydraulics by statistical inference is appealing
because the complete information about phenomena in hydraulics seldom
exists; for example sediment transport, also the more fundamental
problem in fluid mechanics of describing the velocity distribution in
fluids under flow. This latter question shall be approached in a
future contribution. Indeed, the nonlocal terms become specially
important in miniaturized devices at submicronic lengths \cite{a17},
or in the design of stratospheric planes, which fly in rarefied gases
in a density regime between the independent particle description and
the purely continuous description. Another particular problem to it
related is the one of obtaining the structures of shock waves in
fluids for wide ranges of Mach numbers \cite{a18}. Moreover, Burnett
approximation of hydrodynamics has been shown to provide substantial
improvement on many features of the flow occurring in several problems
in hydrodynamics, e.g. the case of Poiseuille flow \cite{a19} and
others \cite{a20}.

The microscopic derivation of a HOH, together with the analysis of the
validity of existing theories, is still a point in question. It has
been shown \cite{a21} that for the case of Maxwellian molecules,
whereas Navier-Stockes approximation yields equations which are stable
against small perturbations, this is not the case when are introduced
Burnett contributions to the equations. It follows that small
perturbations to the solutions, which are periodic in the space
variable with a wavelength smaller than a critical length, are
exponentially unstable. This fact has been called {\it Bobylev's
  instabilitiy}. More recently, Garcia-Colin and collaborators
\cite{a22} have extended Bobylev's analysis for the case of any
interaction potential, and have argued that one can interpret the fact
as to give a bound for a Knudsen number above which the Burnett
equations are not valid. Moreover, Karlin \cite{a23} reconsidered the
question looking for exact solutions to simplified models: when a
linearized ten-moment Grad-method is used, and the Chapman-Enskog
method is applied to the model, in fact there follow instabilities in
the higher-order approximations. On the other hand, resorting to the
Chapman-Enskog solution for linearized Grad ten-moment equations
resummed exactly, solutions are obtained for which the stability of
higher-order hydrodynamics, in various approximations, can be
discussed. 

Furthemore, inclusion of nonlinearity in the theory, in a
Nonlinear Higher-Order Thermo-Hydrodynamics (NLHOTH for short and meaning 
thermal physics of fluid continua), leads 
to additional possible singularities, called hydrodynamic
singularities, as, for example, described in Refs.[24] and [25]. 
A satisfactory construction of a NLHOTH is highly desirable for covering 
a large class of hydrodynamic situations obtaining an understanding of 
the physics involved from the microscopic level, and in the last instance 
gaining insights into technological and industrial processes as in, for 
instance, hydraulic engineering, food engineering, soft-matter engineering, 
etc., which have an associated economic interest. We do present here a 
description of the derivation of a NLHOTH which is based on a formalism 
in Statistical Mechanics shown to be quite appropriate for dealing with 
systems in far-from-equilibrium conditions.
\section{Theoretical Background}
For building a nonlinear higher-order (generalized)
thermo-hydrodynamics on mechanical-statistical basis, one needs a 
nonequilibrium ensemble formalism for open systems. In Ref.[26] has
been described an information-theoretic approach to the construction
of {\it nonequilibrium ensembles}. It involves a variational method
which codifies the derivation of probability distributions - which are
also obtained by heuristic approaches or projection operator
techniques - \cite{a27}, systematizing the work on the subject of a
number of renowned scientists published along the past century.

According to theory, immediatly after the open system of $N$
particles, in contact with external sources and reservoirs, has been
driven out of equilibrium to describe its state requires to introduce
all its observables. But this is equivalent to have acess to the
so-called one-particle (or single-particle), $\hat{n}_{1}$, and
two-particle, $\hat{n}_{2}$, dynamical operators for any subset of the
particles involved: This is so because all usual observable quantities can
be expressed at the microscopic mechanical level in terms of these
operators (e.g. Refs.[28] and [29]), what is described in Appendix A.

On the basis of the construction of the nonequilibrium statistical
operator \cite{a26,a27}, and taking into account the fact that a
complete description of the nonequilibrium state of the system follows
from the knowledge of the single- and two-particle density operators
[or equivalently the density matrix operator, cf. Eqs.(A1) and (A2)],
the most complete statistical distribution is the one built in terms
of the auxiliary (''instantaneously frozen'') statistical operator
\begin{eqnarray}
\bar{{\mathcal R}}(t,0) = \bar{\rho}(t,0) \times \rho_{R} , 
\end{eqnarray}
where
\begin{eqnarray}
\bar{\rho}(t,0) &=& \exp \Bigg\{ - \phi(t) - \sum_{\lambda} 
\int d^{3}r \int d^{3}p F_{1\lambda}({\mathbf r},{\mathbf p};t) \, 
\hat{n}_{1\lambda}({\mathbf r},{\mathbf p}) \nonumber \\
&&- \sum_{\lambda} \int d^{3}r \int d^{3}p \int d^{3}r' \int d^{3}p' 
F_{2\lambda}({\mathbf r},{\mathbf p},{\mathbf r}',{\mathbf p}';t) \, 
\hat{n}_{2\lambda}({\mathbf r},{\mathbf p},{\mathbf r}',{\mathbf p}') 
\Bigg\} ,
\end{eqnarray}
in the clasical case (the quantum one is given in Appendix B). 
Index $\lambda$($=1, 2, \hdots, s$) refers to the possible different
$s$ subsystems of particles (say, different chemical species in a
multicomponent fluid, different classes of quasi-particles in a solid,
namely, electrons in Bloch bands, phonons, etc.); ${\mathbf r}$ and 
${\mathbf p}$ are the so-called position and momentum field
variables. Moreover, we have simplified the matter considering the case of
contact and interaction of the system of interest with ideal reservoir
in stationary states, characterized by the statistical operator 
$\rho_{R}$, so that the complete statistical operator can be
factorized in the form of Eq.(1). Hence, $\bar{\rho}(t,0)$ depends on
the variables of the system of interest and $\rho_{R}$ on the
variables of the reservoir; both distributions are taken as normalized
- as it should -, with $\phi(t)$ ensuring the normalization of 
$\bar{\rho}$, meaning that
\begin{eqnarray}
\phi(t) &=& \int d\Gamma \, \exp \Bigg\{ - \sum_{\lambda} 
\int d^{3}r \int d^{3}p F_{1\lambda}({\mathbf r},{\mathbf p};t) \, 
\hat{n}_{1\lambda}({\mathbf r},{\mathbf p}) \nonumber \\
&&- \sum_{\lambda} \int d^{3}r \int d^{3}p \int d^{3}r' \int d^{3}p' 
F_{2\lambda}({\mathbf r},{\mathbf p},{\mathbf r}',{\mathbf p}';t) \, 
\hat{n}_{2\lambda}({\mathbf r},{\mathbf p},{\mathbf r}',{\mathbf p}') 
\Bigg\} ,
\end{eqnarray}
and $F_{1\lambda}$ and $F_{2\lambda}$ are the intensive nonequilibrium
variables conjugated to $\hat{n}_{1}$ and $\hat{n}_{2}$ (the Lagrange
multiplers in the variational approach). Moreover, $d\Gamma$ is the
element of the volume in the phase space of the system, and for
simplicity we have omitted to indicate the dependence on $\Gamma$ of 
$\hat{n}_{1}$, $\hat{n}_{2}$, $\bar{\rho}$, $\bar{{\mathcal R}}$, and
that $\rho_{R}$ depends on the point phase $\Gamma_{R}$ in the phase
space of the reservoir. 

We recall that $\bar{\rho}$ of Eq.(2) is not the statistical operator
of the nonequilibrium system, but and auxiliary one - as noticed
called the ''instantaneously frozen quasiequilibrium'' statistical
operator -, but which allows to built the proper nonequilibrium
statistical operator (cf. Eq.(4) below), which needs to include {\it
  historicity and irreversibility effects} not present in
$\bar{\rho}$, which does not account for dissipative processes,
besides not providing correct average values in the calculation of
transport coefficients and response functions. 

Finally, the statistical operator explicity written is given by
\begin{eqnarray}
{\mathcal R}_{\varepsilon}(t) = \exp \Bigg\{ \ln \bar{\rho}(t,0) - 
\int_{-\infty}^{t} dt'\, e^{\varepsilon(t'-t)} \frac{d \hfill}{dt'} 
\ln \bar{\rho}(t',t'-t) \Bigg\} \times \rho_{R} ,
\end{eqnarray}
with $\bar{\rho}(t,0)$ of Eq.(2), and we recall that 
\begin{eqnarray}
\bar{\rho}(t',t'-t) = \exp \Big\{ i (t-t') {\mathcal L} \Big\} 
\bar{\rho}(t',0) ,
\end{eqnarray}
(${\mathcal L}$ is the Liouvillian operator of the system meaning 
$i {\mathcal L} \hat{A} = \{\hat{A}, \hat{H} \}$), which is the
auxiliary operator carrying on the mechanical evolution of the system
under Hamiltonian $\hat{H} = \hat{H}_{0} + \hat{H}_{1} + \hat{W} = 
\hat{H}_{0} + \hat{H}'$, where we have introduced $\hat{H}' = 
\hat{H}_{1} + \hat{W}$; with $\hat{H}_{0}$ being the kinetic energy
operator, $\hat{H}_{1}$ contains the internal interactions and $\hat{W}$ 
accounts for the interaction of the system with reservoirs and sources. 
Finally, $\varepsilon$ is an infinitesinal positive real number which is 
taken going to zero after the traces in the calculation of averages 
have been performed (it is present in a kernel that introduces 
irreversibility in the calculations, in a Krylov-Bogoliubov sence). 
We stress that the second contribution in the exponent in Eq.(4) 
accounts for historicity and irreversible behavior from the initial 
time (taken in the remote past, $t_{0} \rightarrow - \infty$, 
implying in adiabatic coupling of correlations, (see for example Ref.[26]), 
or alternatively, can be seen as the adiabatic coupling of the 
interactions responsable for relaxation  processes \cite{a30}). 
Moreover we notice that the time derivative in Eq.(4) takes care of 
the change in time of the thermodynamic state of the system (in the 
first term on the argument i.e. $t'$) and of the microscopic 
mechanical evolution [second term in the argument, i.e. $t'-t$ and see Eq.(6)], 
and that the initial value condition is 
${\mathcal R}_{\varepsilon}(t_{0}) = \bar{\rho}(t_{0},0)$ for $t_{0}
\rightarrow - \infty$.
 
But, as shown elsewhere \cite{a26} it is quite convenient, and
intuitively more satisfactory, to work in an alternative description
than the one in Eq.(2), namely, the so-called {\it generalized
  nonequilibrium grand-canonical ensemble}. In words, it consists into
introducing as basic variables independent linear combinations of
$\hat{n}_{1}$ and $\hat{n}_{2}$. This can be done along several ways
\cite{a27}, but once we are here working within the framework of
classical mechanics, we present a simple derivation in Appendix
C. According to the results presented in the latter, and once for
simplicity we are considering the case of the presence of only one
kind of particles in the system [i.e. $\lambda = 1$ in Eq.(2)], the
auxiliary generalized classical nonequilibrium grand-canonical
statistical operator is given by 
\begin{eqnarray}
\bar{\rho}(t,0) = \exp \left\{ - \hat{S}(t,0) \right\} ,
\end{eqnarray}
where $\hat{S}(t,0)$ is the so-called informational entropy operator 
\cite{a31}, which we write as composed of two contributions, namely
\begin{eqnarray}
\hat{S}(t,0) = \hat{S}_{1}(t,0) + \hat{S}_{2}(t,0) ,
\end{eqnarray}
with $\hat{S}_{1}$ and $\hat{S}_{2}$ given by 
\begin{eqnarray}
\hat{S}_{1}(t,0) &=& \phi_{1}(t) + \int d^{3}r \, \Bigg\{ 
A({\mathbf r},t) \, \hat{n}({\mathbf r}) + 
F_{h}({\mathbf r},t) \, \hat{h}({\mathbf r}) + 
{\mathbf V}({\mathbf r},t) \cdot \hat{{\mathbf I}}_{n}({\mathbf r}) + 
{\mathbf F}_{h}({\mathbf r},t) \cdot \hat{{\mathbf I}}_{h}({\mathbf r}) 
\nonumber \\
&&+ \sum_{\ge 2} \left[ 
F^{[r]}_{h}({\mathbf r},t) \otimes \hat{I}^{[r]}_{h}({\mathbf r}) + 
F^{[r]}_{n}({\mathbf r},t) \otimes \hat{I}^{[r]}_{n}({\mathbf r}) 
\right] \Bigg\} ,
\end{eqnarray}
\begin{eqnarray}
\hat{S}_{2}(t,0) &=& \phi_{2}(t) + \int d^{3}r \int d^{3}r' 
\sum_{p,p'} \sum_{r,r'} F^{[r+r']}_{pp'}({\mathbf r},{\mathbf r}',t) 
\otimes \hat{{\mathcal C}}_{pp'}^{[r+r']}({\mathbf r},{\mathbf r}') ,
\end{eqnarray}
where $p$ and $p'$ stands for indexes $n$ or $h$, and $r$ and $r'$ are
equal to zero (the densities), $1$ (the vectorial fluxes), and 
$2, 3, \hdots$ (higher order fluxes). In the last two equations we
have introduce the definitions 
\begin{eqnarray}
\hat{h}({\mathbf r}) = \int d^{3}p \, \frac{p^{2}}{2m} \, 
\hat{n}_{1}({\mathbf r},{\mathbf p}) , 
\end{eqnarray}
\begin{eqnarray}
\hat{n}({\mathbf r}) = 
\int d^{3}p \, \hat{n}_{1}({\mathbf r},{\mathbf p}) , 
\end{eqnarray}
\begin{eqnarray}
\hat{{\mathcal I}}_{h}({\mathbf r}) = \int d^{3}p \, \frac{p^{2}}{2m} \, 
{\mathbf u} \, \hat{n}_{1}({\mathbf r},{\mathbf p}) ,
\end{eqnarray}
\begin{eqnarray}
\hat{{\mathcal I}}_{n}({\mathbf r}) = \int d^{3}p \,  
{\mathbf u} \, \hat{n}_{1}({\mathbf r},{\mathbf p}) , 
\end{eqnarray}
\begin{eqnarray}
\hat{I}_{h}^{[r]}({\mathbf r}) = \int d^{3}p \, \frac{p^{2}}{2m} \, 
u^{[r]} \, \hat{n}_{1}({\mathbf r},{\mathbf p}) , 
\end{eqnarray}
\begin{eqnarray}
\hat{I}_{n}^{[r]}({\mathbf r}) = \int d^{3}p \, 
u^{[r]} \, \hat{n}_{1}({\mathbf r},{\mathbf p}) ,
\end{eqnarray}
where [cf. Eqs.(C8) and (C9)] 
\begin{eqnarray}
u^{[r]} = [ {\mathbf u} \hdots r-times \hdots {\mathbf u} ] ,
\end{eqnarray}
with $[ \hdots ]$ being the tensorial product of $r$ times the
generating velocity (a vector)
\begin{eqnarray}
{\mathbf u} = {\mathbf p} / m ,
\end{eqnarray}
which is the group velocity of the particle with energy-dispersion
realtion $\varepsilon({\mathbf p}) = p^{2} / 2m$, but the present
formalism also encompasses the case of systems with any kind of
energy-dispersion relation, as, for example, electrons in quantum
states $|{\mathbf k} >$ of Bloch bands (when we take ${\mathbf p} =
\hbar \, {\mathbf k}$), or phonons associated to lattice vibrations
(for acoustic phonons in Debye model $\varepsilon({\mathbf p}) = 
s \, |{\mathbf q}|$ (taking ${\mathbf p} = \hbar \, {\mathbf q}$ with 
${\mathbf q}$ running in Brillouin zone, and where $s$ is the velocity
of propagation), and similarly for photons in black-body radiation
with $\varepsilon({\mathbf p}) = c \, |{\mathbf k}|$ (for 
${\mathbf p} = \hbar \, {\mathbf k}$, and $c$ is the velocity of
light): the general case is briefly summarized in Appendix D. 

The operators in Eqs.(10) and (11) are, respectively, the density of
kinetic energy and of particles, whose integration in space provides
the kinetic energy Hamiltonian and the number of particles. The vector
operators of Eqs.(12) and (13) are the fluxes of the two previous
densities, called first-order fluxes or currents, and those of
Eqs.(14) and (15), with $r \ge 2$, are the higher-order (tensorial)
fluxes of kinetic energy and of matter (or particles). All of them are
present in $\hat{S}_{1}$, while in $\hat{S}_{2}$ we find the tensorial
operators (of rank $r + r'$) 
\begin{eqnarray}
\hat{{\mathcal C}}_{pp'}^{[r+r']}({\mathbf r},{\mathbf r}') = 
\left[ \hat{I}^{[r]}_{k}({\mathbf r}) \, 
\hat{I}^{[r]}_{k'}({\mathbf r}') \right] ,
\end{eqnarray}
where, we recall, $[ \hdots ]$ stands for the tensorial product, $p$
and $p'$ for $h$ or $n$ (referring to energy and particle densities),
and $r$ and $r'$ are $0, 1, 2, \hdots$. 

The average values of the operators of Eqs.(10) to (15) - associated
to the single-particle dynamical operator -, and those of Eq.(18) -
associated to the two-particle dynamical operator -, define the set of
basic macrovariables for the thermo-hydrodynamics, namely
\begin{eqnarray}
I^{[r]}_{p}({\mathbf r},t) = \int d\Gamma \, 
\hat{I}^{[r]}_{p}({\mathbf r}) \, {\mathcal R}_{\epsilon}(t) ,
\end{eqnarray}
for $r = 0$ (the densities), $r = 1$ (the vectorial fluxes or
currents), and $r \ge 2$ (the higher-order fluxes). Hence, we do have 
\begin{eqnarray}
h({\mathbf r},t) = \int d^{3}p \, \frac{p^{2}}{2m} \, 
\left\langle \hat{n}_{1}({\mathbf r},{\mathbf p}) |t \right\rangle , 
\end{eqnarray}
\begin{eqnarray}
n({\mathbf r},t) = \int d^{3}p \, 
\left\langle \hat{n}_{1}({\mathbf r},{\mathbf p}) |t \right\rangle , 
\end{eqnarray}
\begin{eqnarray}
{\mathcal I}_{h}({\mathbf r},t) = \int d^{3}p \, \frac{p^{2}}{2m} \, 
{\mathbf u} \, \left\langle \hat{n}_{1}({\mathbf r},{\mathbf p}) 
|t \right\rangle ,
\end{eqnarray}
\begin{eqnarray}
{\mathcal I}_{n}({\mathbf r},t) = \int d^{3}p \,  
{\mathbf u} \, \left\langle \hat{n}_{1}({\mathbf r},{\mathbf p}) 
|t \right\rangle , 
\end{eqnarray}
\begin{eqnarray}
I_{h}^{[r]}({\mathbf r},t) = \int d^{3}p \, \frac{p^{2}}{2m} \, 
u^{[r]} \, \left\langle \hat{n}_{1}({\mathbf r},{\mathbf p}) 
|t \right\rangle , 
\end{eqnarray}
\begin{eqnarray}
I_{n}^{[r]}({\mathbf r},t) = \int d^{3}p \, 
u^{[r]} \, \left\langle \hat{n}_{1}({\mathbf r},{\mathbf p}) 
|t \right\rangle ,
\end{eqnarray}
where
\begin{eqnarray}
\left\langle \hat{n}_{1}({\mathbf r},{\mathbf p}) |t \right\rangle = 
\int d\Gamma \, \hat{n}_{1}({\mathbf r},{\mathbf p}) \, 
{\mathcal R}_{\epsilon}(t) .
\end{eqnarray}

Similarly for the contributions associated to the two-particle
dynamical operator we do have the corresponding thermodynamic
variables 
\begin{eqnarray}
{\mathcal C}_{pp'}^{[r+r']}({\mathbf r},{\mathbf r}',t) = 
\int d\Gamma \, 
\hat{{\mathcal C}}_{pp'}^{[r+r']}({\mathbf r},{\mathbf r}') \, 
{\mathcal R}_{\epsilon}(t) ,
\end{eqnarray}
with ${\mathcal C}$ given by Eq.(18).

The complete set of thermo-hydrodynamic variables are presented in
Table I organized, as one is going up in both columns, in increasing
order of tensorial rank.
\begin{table}[h]
\begin{tabular}{||c|c||}\hline 
\multicolumn{2}{||c||}{Generalized Grand-Canonical Description} \\ \hline 
Single-Particle &Two-Particle \\ \hline
$\hdots$ &$\hdots$ \\ \hline 
$I_{h}^{[r]}({\mathbf r},t)$~; $I_{n}^{[r]}({\mathbf r},t)$ &
${\mathcal C}_{pp'}^{[r_{1}+r_{2}]}({\mathbf r},{\mathbf r}',t)$~; 
$r_{1} + r_{2} = r$ \\ \hline
$\hdots$ &$\hdots$ \\ \hline
$I_{h}^{[2]}({\mathbf r},t)$~; $I_{n}^{[2]}({\mathbf r},t)$ &
$\: \: \: \: \: \: \: \: \:$ 
${\mathcal C}_{pp'}^{[1+1]}({\mathbf r},{\mathbf r}',t)$~; 
${\mathcal C}_{pp'}^{[2+0]}({\mathbf r},{\mathbf r}',t)$~; 
${\mathcal C}_{pp'}^{[0+2]}({\mathbf r},{\mathbf r}',t)$ 
$\: \: \: \: \: \: \: \: \:$ \\ \hline
${\mathbf I}_{h}({\mathbf r},t)$~; ${\mathbf I}_{n}({\mathbf r},t)$ &
${\mathcal C}_{pp'}^{[1+0]}({\mathbf r},{\mathbf r}',t)$~; 
${\mathcal C}_{pp'}^{[0+1]}({\mathbf r},{\mathbf r}',t)$~; \\ \hline
$h({\mathbf r},t)$~; $n({\mathbf r},t)$ &
${\mathcal C}_{pp'}^{[0+0]}({\mathbf r},{\mathbf r}',t)$\\ \hline
$\: \: \: \: \: \: \: \: \:$ Dynamical Density Operator 
$\: \: \: \: \: \: \: \: \:$&
Dynamical Density Operator \\ 
$\hat{n}_{1}({\mathbf r},{\mathbf p})$ &
$\hat{n}_{2}({\mathbf r},{\mathbf p},{\mathbf r}',{\mathbf p}')$ \\
\hline
\end{tabular}
\caption{The basic set of macrovariables one- and two-particle
dynamical operators description and the generalized grand-canonical 
description.}
\end{table}

We call the attention to the fact that the {\it generalized 
nonequilibrium grand-canonical statistical operator} is the one of
Eq.(4) once in it is introduced the auxiliary operator of Eq.(7);
explicitly 
\begin{eqnarray}
{\mathcal R}_{\epsilon}(t) = \rho_{\epsilon}(t) \times \rho_{R} ,
\end{eqnarray}
with 
\begin{eqnarray}
\rho_{\epsilon}(t) = \exp \left\{ - \hat{S}(t,0) + \int_{-\infty}^{t}
dt'\, e^{\epsilon (t'-t)} \, \frac{d \hfill}{dt'} \hat{S}(t,t'-t) 
\right\} . 
\end{eqnarray}

Furthemore, we recall that it can be introduced the separation of 
$\rho_{\epsilon}$ into two parts [coming from each contribution in the
exponent of Eq.(20)], namely 
\begin{eqnarray}
\rho_{\epsilon}(t) = \bar{\rho}(t,0) + \rho_{\epsilon}^{\prime}(t) , 
\end{eqnarray}
with $\bar{\rho}$ of Eq.(6), and that for the basic variables (in this
case those of Table I) and only the basic variables, it follows that
\begin{eqnarray}
I_{p}^{[r]}({\mathbf r},t) = \int d\Gamma \, 
\hat{I}_{p}^{[r]}({\mathbf r}) \, \rho_{\epsilon}(t) = 
\int d\Gamma \, \hat{I}_{p}^{[r]}({\mathbf r}) \, \bar{\rho}(t,0) ,
\end{eqnarray}
\begin{eqnarray}
{\mathcal C}_{pp'}^{[r+r']}({\mathbf r},{\mathbf r}';t) = 
\int d\Gamma \, 
\hat{{\mathcal C}}_{pp'}^{[r+r']}({\mathbf r},{\mathbf r}') \, 
\rho_{\epsilon}(t) =  \int d\Gamma \, 
\hat{{\mathcal C}}_{pp'}^{[r+r']}({\mathbf r},{\mathbf r}') \, 
\bar{\rho}(t,0) ,
\end{eqnarray}
i.e., $\rho_{\epsilon}^{\prime}$ does not contribute for the average
value of, we stress, the basic variables only \cite{a4,a26}. Finally, 
it can be noticed that the quantities of Eq.(32) can be related to
fluctuations of the densities and their fluxes. In fact we can write 
\begin{eqnarray}
\sigma_{pp'}^{[r+r']}({\mathbf r},{\mathbf r}',t) &=& 
\int d\Gamma \, \left( 
\hat{I}_{p}^{[r]}({\mathbf r}) - I_{p}^{[r]}({\mathbf r},t) \right) 
\left( \hat{I}_{p'}^{[r']}({\mathbf r}') - 
I_{p'}^{[r']}({\mathbf r}',t) \right) \, \rho_{\epsilon}(t) 
\nonumber \\
&=& {\mathcal C}_{pp'}^{[r+r']}({\mathbf r},{\mathbf r}';t) - 
I_{p}^{[r]}({\mathbf r},t) \, I_{p'}^{[r']}({\mathbf r}',t) ,
\end{eqnarray}
which are instantaneous at time $t$ but containing space correlations.

Let us next proceed to derive the equations of evolution of the basic
variables, that is, the generalized hydrodynamic equations.
\section{Nonlinear Higher-Order Thermo-Hydrodynamics (NLHOTH)}
We begin considering a NLHOTH for a system of independent particles,
say, an ideal fluid or, more generally, a system of particles with the
interaction between them treated in an average-field approximation (a
quite interesting case is the fluid of mobile electrons in
crystals). Hence, the set of basic variables consists of the one on
the left side of Table I, and Eq.(19), that is the one composed of 
\begin{eqnarray}
\left\{ h({\mathbf r},t), n({\mathbf r},t), 
{\mathbf I}_{h}({\mathbf r},t), {\mathbf I}_{n}({\mathbf r},t), 
\left\{ I_{h}^{[r]}({\mathbf r},t) \right\}, 
\left\{ I_{h}^{[r]}({\mathbf r},t) \right\} \right\} .
\end{eqnarray}
The hydrodynamical motion of this fluid of single particles is
described by the set of coupled highly-nonlinear integro-differential
equations consisting of the kinetic equations provided by the
formalism (see Appendix E), which are the {\it generalized hydrodynamic
equations} given by 
\begin{eqnarray}
\frac{\partial \hfill}{\partial t} h({\mathbf r},t) + 
\nabla \cdot {\mathbf I}_{h}({\mathbf r},t) = 
J_{h}({\mathbf r},t) ,
\end{eqnarray}
\begin{eqnarray}
\frac{\partial \hfill}{\partial t} n({\mathbf r},t) + 
\nabla \cdot {\mathbf I}_{n}({\mathbf r},t) = 
J_{n}({\mathbf r},t) ,
\end{eqnarray}
\begin{eqnarray}
\frac{\partial \hfill}{\partial t} {\mathbf I}_{h}({\mathbf r},t) + 
\nabla \cdot I_{h}^{[2]}({\mathbf r},t) = 
{\mathbf J}_{h}({\mathbf r},t) ,
\end{eqnarray}
\begin{eqnarray}
\frac{\partial \hfill}{\partial t} {\mathbf I}_{n}({\mathbf r},t) + 
\nabla \cdot I_{n}^{[2]}({\mathbf r},t) = 
{\mathbf J}_{n}({\mathbf r},t) ,
\end{eqnarray}
\begin{eqnarray}
\frac{\partial \hfill}{\partial t} I_{h}^{[r]}({\mathbf r},t) + 
\nabla \cdot I_{h}^{[r+1]}({\mathbf r},t) = 
J_{h}^{[r]}({\mathbf r},t) ,
\end{eqnarray}
\begin{eqnarray}
\frac{\partial \hfill}{\partial t} I_{n}^{[r]}({\mathbf r},t) + 
\nabla \cdot I_{n}^{[r+1]}({\mathbf r},t) = 
J_{n}^{[r]}({\mathbf r},t) ,
\end{eqnarray}
with $r \ge 2$, $\nabla \cdot$ is the tensorial divergence operator,
and where, on the right, are present the collision operators
\begin{eqnarray}
J_{p}^{[r]}({\mathbf r},t) = \int_{-\infty}^{t} dt' \, 
e^{\epsilon (t'-t)} \, \int d\Gamma \left\{ \left\{ 
\hat{I}_{p}^{[r]} , \hat{H}^{\prime}(t'-t)_{0} \right\}, 
\hat{H}^{\prime} \right\} \, \rho_{\epsilon}^{\prime}(t) \times 
\rho_{R} ,
\end{eqnarray}
for $r = 0, 1, 2, \hdots$, $p = h$ or $n$, $\left\{ \hdots \right\}$
is Poisson parentheses (see Appendix E), and subindex nought stands
for evolution in the interaction representation. We recall that 
$\hat{H}^{\prime}$ contains the internal interactions plus the one
with the reservoir, and, eventually, with external sources of
perturbation. Moreover, if the collision operators are neglected we do
have the conservation equation for each quantity. As a matter of
practicality, one needs - depending on the problem in hands - to
resort to a {\it truncation} in the chain of generalized hydrodynamic
equations, Eqs.(35) to (40), say for $r = n$: the point is discussed
in Ref.[26], and further considerations are presented in
Refs.[11,27]. Hence, once the equation for the $n$-th flux contains a
term with the divergence of the $(n+1)$-th flux, to close the system
of hydrodynamic variables one needs to express the latter in terms of 
all the other basic variables (the densities and the fluxes up to
order $n$), and this involves a nonlinear expression. Thus we do face
nonlinearities at this point, but also, and more importantly, in the
collision integrals $J$. They are - see Eq.(41) - depending on a
highly nonlinear way on the nonequilibrium thermodynamic variables
$F$'s [cf. Eqs.(8) and (9)], and therefore the system of hydrodynamic
equations needs be coupled with the set of nonequilibrium equations of
states, namely, Eqs.(20) to (25) and (27) (let us noticed that such
dependence comes from the statistical operator in the calculation of
averages). Consequently, the collision operators depend on a highly
nonlinear way on the nonequilibrium thermodynamic variables, which, in
turn depend on a highly nonlinear way on the basic hydrodynamic
variables. All this is illustrated in the follow up article in a
simple model and a first-order hydrodynamics ($n = 1$).

Summarizing, Eqs.(35) to (40) provide the thermo-hydrodynamics of the
nonequilibrium many-body system, which describe the evolution of an 
infinite set of coupled highly-nonlinear integro-differential
equations. Moreover, as noticed, practical usage requires us to
introduce a truncation procedure, meaning the analog of the one in the
Hilbert-Chapmann-Enskog approach to the Boltzmann equation, or, more
precisely, one which is closely related to Grad's moments method
\cite{a32}: This is discussed in Ref.[33], and a particular case - a
phoinjected plasma in semiconductors - is considered in Refs.[34] and
[35]; we return to this important question in a future article on
NLHOTH. 

It can also be noticed that, alternatively, Eqs.(35) to (40) can be
closed writting on the right-hand side the basic variables in terms of
the intensive ones, but for building this NLHOTH it is more
satisfactory the first choice of description (i.e. in term of the
variables of Eq.(34)). 

We also call the attention to the fact that if we introduce a
nonequilibrium generalized grand-canonical-like partition function as 
\begin{eqnarray}
\phi(t) = \ln \bar{{\mathcal Z}}(t) ,
\end{eqnarray}
with $\phi(t)$ of Eq.(3) but for $F_{2} = 0$ once we are working in a
single-particle representation, it follows that
\begin{eqnarray}
I_{p}^{[r]}({\mathbf r},t) = - \delta \ln \bar{{\mathcal Z}}(t) / 
\delta F_{p}^{[r]}({\mathbf r},t) ,
\end{eqnarray}
an alternative of Eq.(23) and where $\delta$ stands for functional
derivative \cite{a36}. Moreover, defining the
nonequilibrium-informational entropy \cite{a37}
\begin{eqnarray}
\bar{S}(t) = \int d\Gamma \, \hat{S}(t,0) ,
\end{eqnarray}
it is verified that
\begin{eqnarray}
F_{p}^{[r]}({\mathbf r},t) = \delta \bar{S}(t) / 
\delta I_{p}^{[r]}({\mathbf r},t) .
\end{eqnarray}

It can be noticed that Eq.(45) and Eq.(43) - the latter an alternative
form of the Eqs.(20) to (25) - relate the basic variables (the
densities and their fluxes) $I_{p}^{[r]}({\mathbf r},t)$, to the 
nonequilibrium thermodynamic variables, $F_{p}^{[r]}({\mathbf r},t)$,
which, as noticed, can be labelled, in analogy to the case of systems
in equilibrium, {\it nonequilibrium equations of state}. As also
commented above, they are fundamental for closing the generalized
hydrodynamic equations, Eqs.(35) to (40), that is, the set of
equations of evolution is coupled to the set of nonequilibrium
equations of state.

On the other hand, if the physical situation we are considering
requires to further introduce as basic variables the
pair-correlation-like contributions of Eq.(27), use must be made of
the statistical operator of Eq.(4) where the auxiliary statistical
operator of Eqs.(6) to (9) must be introduced. The set of Eqs.(35) to
(40) - in which it must be understand that the average values are
calculated using the above said statistical operator -, is now
extended to incorporate the equations of evolution of the added
variables, namely
\begin{eqnarray}
\frac{\partial \hfill}{\partial t}
{\mathcal C}_{pp'}^{[r+r']}({\mathbf r},{\mathbf r}';t) &=& 
\int d\Gamma \left\{ 
\hat{{\mathcal C}}_{pp'}^{[r+r']}({\mathbf r},{\mathbf r}') ,
\hat{H}_{0} \right\} \, \bar{\rho}(t,0) \nonumber \\
&&+ 
\int d\Gamma \int d\Gamma_{R} \, \left\{ 
\hat{{\mathcal C}}_{pp'}^{[r+r']}({\mathbf r},{\mathbf r}') ,
\hat{H}^{\prime} \right\} \, \rho^{\prime}_{\epsilon}(t) \times
\rho_{R} , 
\end{eqnarray}
or 
\begin{eqnarray}
&&\frac{\partial \hfill}{\partial t}
{\mathcal C}_{pp'}^{[r+r']}({\mathbf r},{\mathbf r}';t) +  
\int d\Gamma \left[ \nabla \cdot \hat{I}_{p}^{r+1]}({\mathbf r}) \, 
\hat{I}_{p'}^{r']}({\mathbf r}') + 
\hat{I}_{p}^{r]}({\mathbf r}) \, \nabla^{\prime} \cdot 
\hat{I}_{p'}^{r'+1]}({\mathbf r}') \right] \, \bar{\rho}(t,0) 
\nonumber \\
&=& \frac{\partial \hfill}{\partial t}
{\mathcal C}_{pp'}^{[r+r']}({\mathbf r},{\mathbf r}';t) + 
\nabla \cdot \int d\Gamma \, 
{\mathcal C}_{pp'}^{[(r+1)+r']}({\mathbf r},{\mathbf r}';t) + 
\nabla^{\prime} \cdot \int d\Gamma \, 
{\mathcal C}_{pp'}^{[r+(r'+1)]}({\mathbf r},{\mathbf r}';t) 
\nonumber \\
&=& {\mathcal J}_{pp'}^{[r+r']}({\mathbf r},{\mathbf r}';t) ,
\end{eqnarray}

The set of equations of evolution needs, as noticed, be coupled to the
equations of state, that is, those of Eqs.(43) or (45), and now
with the ones associated to the two-particle contributions, namely
\begin{eqnarray}
{\mathcal C}_{pp'}^{[r+r']}({\mathbf r},{\mathbf r}';t) = 
\delta \ln \bar{{\mathcal Z}}(t) / 
\delta F_{pp'}^{[r+r']}({\mathbf r},{\mathbf r}';t) , 
\end{eqnarray}
or
\begin{eqnarray}
\delta F_{pp'}^{[r+r']}({\mathbf r},{\mathbf r}';t) = 
- \delta S(t) / 
\delta {\mathcal C}_{pp'}^{[r+r']}({\mathbf r},{\mathbf r}';t) .
\end{eqnarray}
We restate that both cases, the one involving only single-particle
operators and the other with single- and two-particle operators, the
{\it nonlinearity} is present in the scattering operators ${\mathcal
  J}$'s, which are in general depending on a highly nonlinear
variables $F$'s, and then, through the nonequilibrium equations of
state, depend on a highly nonlinear way on the basic variables. This
is illustrated in the follow-up article. 
\section{Concluding Remarks}
Briefly summarizing the results, it has been shown how a statistical
nonequilibrium ensemble formalism (applicable to the study of systems
even in conditions far from equilibrium) provides - via the use of a 
quite general nonequilibrium grand-canonical ensemble - a microscopic
foundation for a Nonlinear Higher-Order and Fluctuations-dependent
Hydrodynamics. Its description is based on the set of macrovariables
consisting of the densities of energy and matter (particles), their
fluxes of all order, and the direct and cross correlations between all
of them: In that way higher-order descriptions and fluctuations,
respectively, are included in the theory, (cf. Table I).

Al these macrovariables are the average value over the nonequilibrium
ensemble of the corresponding microscopic mechanical operators. Once
the complete set of macrovariables is given we can obtain the
nonlinear hydrodynamic equations, which are the average value over the
nonequilibrium enesemble of Hamilton equations of motion (in the
classical level or Heisenberg equations at the quantum level) of the
basic microvariables (mechanical observables): cf. Eqs.(35) to (40) and
(46) to (47). 

These generalized hydrodynamic equations present on the left side the
conserving part of the corresponding quantity, and on the right-hand
side are present the collision integrals which include the action of
external sources and the contributions of scattering processes
responsible for dissipative effects. As noticed before, these
collision integrals depend on a highly nonlinear way on the
nonequilibrium thermodynamic variables (the Lagrange multipliers that
the variational method introduces, namely, $F_{p}^{[r]}$ and
$F_{pp'}^{[r+r']}$ in Eqs.(7) to (9)). Hence, this set of hydrodynamic
equations is closed after it is coupled with the set of nonequilibrium
equations of state, that is, Eqs.(43) and (48), or Eqs.(45) and (49)
or simply Eqs.(19) and (27). 

In that way we do have a quite generalized hydrodynamics under any
arbitrary condition of excitation, which, as noticed, can be referred
to as Nonlinear Higher-Order and correlation-dependent Hydrodynamics. In
the follow up article we present an illustration of its application,
consisting of a simple model of two ideal fluids in mutual interaction
- one of them acting as a reservoir - and in a truncated NLHOH of
order 1. 
\begin{acknowledgments}
We acknowledge financial support from S\~ao Paulo State Research 
Foundation (FAPESP). ARV and RL are Brazil National Research Council 
(CNPq) research fellows.
\end{acknowledgments}
\appendix
\section{Dynamical Density Operators (Reduced Distribution Functions)}
For the sake of completeness, we notice that in classical mechanics
the one-particle and two-particle operators $\hat{n}_{1}$ and 
$\hat{n}_{2}$ are given, respectively, by 
\begin{eqnarray}
\hat{n}_{1}({\mathbf r},{\mathbf p}) = 
\sum_{j=1}^{N} \delta( {\mathbf r} - {\mathbf r}_{j} ) \, 
\delta( {\mathbf p} - {\mathbf p}_{j} ) ,
\end{eqnarray}
\begin{eqnarray}
\hat{n}_{2}({\mathbf r},{\mathbf p},{\mathbf r}',{\mathbf p}') = 
\sum_{j=1}^{N} \sum_{k=1}^{N} \delta( {\mathbf r} - {\mathbf r}_{j} ) \, 
\delta( {\mathbf p} - {\mathbf p}_{j} ) \, 
\delta( {\mathbf r}' - {\mathbf r}_{k} ) \, 
\delta( {\mathbf p}' - {\mathbf p}_{k} ) ,
\end{eqnarray}
for each one of the different $S$ types of particles that can be
present [cf. Eq.(2)], and where, as usual, ${\mathbf r}_{j}$ and 
${\mathbf p}_{j}$ are the coordinate and momentum of the $j$-th
particle, and ${\mathbf r}$ and ${\mathbf p}$ the so-called coordinate
and momentum field variables.

In quantum mechanics the one- and two-particle density operators are
($\sigma$ is the spin index)
\begin{eqnarray}
\hat{n}_{1}({\mathbf r},\sigma,{\mathbf r}',\sigma') &=& 
\Psi_{\sigma}^{\dagger}({\mathbf r}) \, 
\Psi_{\sigma}^{\dagger}({\mathbf r}) , \\[12pt]
\hat{n}_{2}({\mathbf r}_{1},\sigma_{1},{\mathbf r}_{2},\sigma_{2},
{\mathbf r}_{2}',\sigma_{2}',{\mathbf r}_{1}',\sigma_{1}') &=& 
\Psi_{\sigma_{1}}^{\dagger}({\mathbf r}_{1}) \, 
\Psi_{\sigma_{2}}^{\dagger}({\mathbf r}_{2}) \, 
\Psi_{\sigma_{2}'}({\mathbf r}_{2}') 
\Psi_{\sigma_{1}'}({\mathbf r}_{1}') , 
\end{eqnarray}
where $\Psi(\Psi^{\dagger})$ are single-particle field operators in
second quantization (an excellent didactical description of them is
available in the article by B. Robertson of Ref.[38]). These operators
can be expressed in terms of annihilation and creation operators in
single-particle states with, say, wavefunction 
$\varphi_{{\mathbf k}\sigma}$ and energy levels 
$\varepsilon_{{\mathbf k}\sigma}$, in the form
\begin{eqnarray}
\Psi_{\sigma}({\mathbf r}) = 
\sum_{{\mathbf k}} \varphi_{{\mathbf k}\sigma} \, c_{{\mathbf k}\sigma} ,
\end{eqnarray}
\begin{eqnarray}
\Psi_{\sigma}^{\dagger}({\mathbf r}) = 
\sum_{{\mathbf k}} \varphi_{{\mathbf k}\sigma}^{\ast} \, 
c_{{\mathbf k}\sigma}^{\dagger} ,
\end{eqnarray}

This is quite convenient since it introduces a description in the
reciprocal ${\mathbf k}$-space, with practical advantages in
calculations and physical interpretations. In this way we can
introduce instead of $\hat{n}_{1}$ and $\hat{n}_{2}$ of Eqs.(A3) and
(A4), matrices composed with the quantities
\begin{eqnarray}
\Bigg\{ 
\left\{ 
c_{{\mathbf k}\sigma}^{\dagger} \, c_{{\mathbf k}'\sigma'} \right\}, 
\left\{ 
c_{{\mathbf k}\sigma}^{\dagger} \, c_{{\mathbf k}'\sigma'}^{\dagger}
\,  c_{{\mathbf k}\sigma} \, c_{{\mathbf k}'\sigma'} \right\} 
\Bigg\} ,
\end{eqnarray}
Hence, any one-particle mechanical observable can be expressed in
terms of the quantities 
\begin{eqnarray}
\left\{ 
\hat{n}_{{\mathbf k}\sigma\sigma'} =  
c_{{\mathbf k}\sigma}^{\dagger} \, c_{{\mathbf k}'\sigma'}, \,  
\hat{n}_{{\mathbf k}{\mathbf Q}\sigma\sigma'} =  
c_{{\mathbf k}+\frac{1}{2}{\mathbf Q}\sigma}^{\dagger} \, 
c_{{\mathbf k}-\frac{1}{2}{\mathbf Q}\sigma} \right\} ,
\end{eqnarray}
with ${\mathbf Q} \neq {\mathbf 0}$, where we have separated out
diagonal terms and non-diagonal ones: the first are so-called
populations and the others coherences \cite{a39} (while the former
appear in the description of whole properties - or uniform states -,
the latter describe local properties - or inhomogeneous states). Any
one-particle mechanical observable is then written in second
quantization in the known form 
\begin{eqnarray}
\hat{A} = \sum_{{\mathbf k}\sigma,{\mathbf k}'\sigma'} 
\left\langle {\mathbf k}\sigma | \hat{A} | {\mathbf k}'\sigma' 
\right\rangle \, 
c_{{\mathbf k}\sigma}^{\dagger} \, c_{{\mathbf k}'\sigma'} ,
\end{eqnarray}
for any pair ${\mathbf k}\sigma$ and ${\mathbf k}'\sigma'$, and where
is present the matrix element of $\hat{A}$ between one-particle states
$\Big\langle {\mathbf k}\sigma |$ and 
$| {\mathbf k}'\sigma' \Big\rangle$. 

For two-particle mechanical observables, using in Eq.(A7) the change
of indexes ${\mathbf k}+\frac{1}{2}{\mathbf Q}$ for 
${\mathbf k}$, ${\mathbf k}'-\frac{1}{2}{\mathbf Q}'$ for 
${\mathbf k}'$, ${\mathbf k}-\frac{1}{2}{\mathbf Q}$ for 
${\mathbf k}_{1}$, ${\mathbf k}'+\frac{1}{2}{\mathbf Q}'$ for 
${\mathbf k}_{2}$, we have 
\begin{eqnarray}
\hat{n}_{{\mathbf k}\sigma_{1},{\mathbf k}'\sigma_{1}',
{\mathbf k}',\sigma_{2},{\mathbf rk},\sigma_{2}'} = 
c_{{\mathbf k}\sigma_{1}}^{\dagger} \,
c_{{\mathbf k}'\sigma_{1}'}^{\dagger} \,  
c_{{\mathbf k}'\sigma_{2}} \,
c_{{\mathbf k}\sigma_{2}'} , 
\end{eqnarray}
for ${\mathbf Q}, {\mathbf Q}' \neq {\mathbf 0}$. A two-particle
observable takes the form \cite{a40,a41} 
\begin{eqnarray}
\hat{B} = \sum_{{\mathbf k},{\mathbf k}',{\mathbf Q},{\mathbf Q}'}
\sum_{\sigma_{1}\sigma_{1}'\sigma_{2}\sigma_{2}'} 
&&\left\langle {\mathbf k}+\frac{1}{2}{\mathbf Q}\sigma_{1}, 
{\mathbf k}'-\frac{1}{2}{\mathbf Q}'\sigma_{2} |  
\hat{B} 
| {\mathbf k}'+\frac{1}{2}{\mathbf Q}'\sigma_{2}', 
{\mathbf k}-\frac{1}{2}{\mathbf Q}\sigma_{1}' \right\rangle \times 
\nonumber \\
&&\times 
\left[ \hat{n}_{{\mathbf k}{\mathbf Q}\sigma_{1}\sigma_{1}'} \, 
\hat{n}_{{\mathbf k}'{\mathbf Q}'\sigma_{2}\sigma_{2}'}^{\dagger}
 \mp \delta_{{\mathbf k}{\mathbf Q}\sigma_{1}\sigma_{1}', 
{\mathbf k}'{\mathbf Q}'\sigma_{2}\sigma_{2}'} \, 
c_{{\mathbf k}+\frac{1}{2}{\mathbf Q}\sigma_{1}}^{\dagger} \,
c_{{\mathbf k}'+\frac{1}{2}{\mathbf Q}'\sigma_{1}'} \right] ,
\end{eqnarray}
where we have rearranged the order in the product of four operators in
Eq.(A11), to obtain a contribution as a product of two
$\hat{n}_{1}$-type operators and amother involving $\hat{n}_{1}$
already present in Eq.(A8); upper sign minus stands for the case of
fermions and lower plus for bosons.

For example, Coulomb interaction between pairs of electrons is given
by 
\begin{eqnarray}
\hat{H}_{C} = \sum_{{\mathbf Q}\neq{\mathbf 0}} {\mathcal V}({\mathbf Q}) 
\left[ 
\hat{n}_{{\mathbf Q}} \, \hat{n}_{{\mathbf Q}}^{\dagger} - N 
\right] , 
\end{eqnarray}
where, once a complete set of plane-wave states has been taken as
one-particle quantum states, ${\mathcal V}({\mathbf Q}) = 4 \pi e^{2} / V
Q^{2}$ ($V$ is the sample volume), 
\begin{eqnarray}
\hat{n}_{{\mathbf Q}} = \sum_{{\mathbf k}\sigma\sigma'} 
\delta_{\sigma\sigma'} \, 
\hat{n}_{{\mathbf k}{\mathbf Q}\sigma\sigma'} ,
\end{eqnarray}
is the ${\mathbf Q}$-wave-number Fourier transform of the operator
density of charge of the electrons (in units of the electron charge),
and $\hat{N} = \sum_{{\mathbf k}\sigma} 
c_{{\mathbf k}\sigma}^{\dagger} \, c_{{\mathbf k}\sigma}$ is the
  operator number of particles. 

We call attention to the fact that if the particles are bosons, then
coherent states are possible to be present and then, besides the one-
and two-particle density operators, we must also add the individuals
amplitudes $c_{{\mathbf k}\sigma}$ (and 
$c_{{\mathbf k}\sigma}^{\dagger}$) for the description of the boson
system (for example coherent photons in a laser, coherent phonons in
semiconductors, etc). 
\section{The Quantum Mechanical Nonequilibrium Statistical Operator}
In the quantum case the auxiliary statistical operator of Eq.(2) is 
\begin{eqnarray}
\bar{\rho}(t,0) = \exp \left\{ - \phi(t) - 
\sum_{{\mathbf k}} F_{{\mathbf k}}(t) \, \hat{n}_{{\mathbf k}} - 
\sum_{{\mathbf k}{\mathbf Q}} \sum_{{\mathbf k}'{\mathbf Q}'} 
F_{{\mathbf k}{\mathbf Q},{\mathbf k}'{\mathbf Q}'}(t) \, 
\hat{n}_{{\mathbf k}{\mathbf Q}} \, 
\hat{n}_{{\mathbf k}'{\mathbf Q}'}^{\dagger} \right\} ,
\end{eqnarray}
which is written in terms of the quantities of Eqs.(A8) and (A11),
with spin indexes omitted. The quantum nonequilibrium statistical
operator is given by Eq.(4), where now enters $\bar{\rho}$ of Eq.(6),
and 
\begin{eqnarray}
\bar{\rho}(t,t'-t) = \exp \left\{ 
- \frac{1}{{\textrm i} \, \hbar} (t'- t) \, \hat{H} \right\} \, 
\bar{\rho}(t,0) 
\exp \left\{ \frac{1}{{\textrm i} \, \hbar} (t'- t) \, \hat{H}
\right\} .
\end{eqnarray}
The three $F$'s in Eq(B1) are the corresponding Lagrange multipliers
in the variational approach (intensive nonequilibrium thermodynamic
variables conjugated to the dynamical operators). Moreover, the
macrovariables for describing the nonequilibrium thermodynamic state
of the system are the average value of the same quantities over the
nonequilibrium ensemble, namely
\begin{eqnarray}
n_{{\mathbf k}}(t) = Tr \left\{ \hat{n}_{{\mathbf k}} \, 
{\mathcal R}_{\epsilon}(t) \right\} , 
\end{eqnarray}
\begin{eqnarray}
n_{{\mathbf k}{\mathbf Q}}(t) = Tr \left\{ 
\hat{n}_{{\mathbf k}{\mathbf Q}} \, 
{\mathcal R}_{\epsilon}(t) \right\} , 
\end{eqnarray}
with ${\mathbf Q} \neq {\mathbf 0}$, and 
\begin{eqnarray}
n_{{\mathbf k}{\mathbf Q},{\mathbf k}'{\mathbf Q}'}(t) = 
Tr \left\{ \hat{n}_{{\mathbf k}{\mathbf Q}} \, 
\hat{n}_{{\mathbf k}'{\mathbf Q}'}^{\dagger} \, 
{\mathcal R}_{\epsilon}(t) \right\} , 
\end{eqnarray}
with no restriction on ${\mathbf Q}$ and ${\mathbf Q}'$.
\section{Classical Nonequilibrium Grand-Canonical Statistical Operator}
Consider thermodynamic intensive variable 
$F_{1}({\mathbf r},{\mathbf p};t)$ in the auxiliary statistical
operator of Eq.(2), and let us perform a Taylor series expansion in 
${\mathbf p}$, i.e.
\begin{eqnarray}
F_{1}({\mathbf r},{\mathbf p};t) &=& \Phi({\mathbf r},t) + 
{\mathbf \Phi}({\mathbf r},t) \cdot {\mathbf p} + 
\Phi^{[2]}({\mathbf r},t) \otimes [ {\mathbf p} {\mathbf p} ]
\nonumber \\
&&+ \sum_{r\geq3} \Phi^{[r]}({\mathbf r},t) \otimes 
[ {\mathbf p} \hdots r-times \hdots {\mathbf p} ] ,
\end{eqnarray}
where 
$\Phi = F_{1}({\mathbf r},{\mathbf p};t)|_{{\mathbf p}={\mathbf 0}}$, 
$\Phi^{[r]} = \frac{1}{r!} 
[ \nabla_{{\mathbf p}} \hdots r-times \hdots \nabla_{{\mathbf p}} ] 
F_{1}({\mathbf r},{\mathbf p};t)|_{{\mathbf p}={\mathbf 0}}$, for 
$r = 1, 2, 3, \hdots$ indicating the tensorial rank, $[ \hdots ]$
stands for tensorial product of vectors, dot as usual as scalar
product, and $\otimes$ for fully contracted product of tensors. Next,
we redefine quantities $\Phi$ in the following way
\begin{eqnarray}
&&\Phi({\mathbf r},t) \equiv A_{n}({\mathbf r},t) , \\[12pt]
&&{\mathbf \Phi}({\mathbf r},t) \equiv 
{\mathbf {\mathcal V}}_{n}({\mathbf r},t) , \\[12pt]
&&\Phi^{[r]}({\mathbf r},t) \equiv 
\frac{F_{n}^{[r]}({\mathbf r},t)}{m^{r}} + 
\frac{1}{3} \left[ \left( \Phi^{[r]}({\mathbf r},t) \otimes 1^{[2]} 
\right) \, 1^{[2]} \right] ,  
\end{eqnarray}
further introducing 
\begin{eqnarray}
\frac{F_{R}^{[r-2]}({\mathbf r},t)}{2 \, m^{r-1}} \equiv 
\frac{1}{3} \Phi^{[r]}({\mathbf r},t) \otimes 1^{[2]} , 
\end{eqnarray}
where $1^{[2]}$ is the unit tensor of rank two, it can be noticed that
it is satisfied the rule
\begin{eqnarray}
F_{n}^{[r]}({\mathbf r},t) \otimes 1^{[2]} = 0 ,
\end{eqnarray}
and that
\begin{eqnarray}
Phi_{n}^{[r]}({\mathbf r},t) \otimes [ {\mathbf p} \hdots r-times \hdots
{\mathbf p} ] = 
F_{n}^{[r]}({\mathbf r},t) \otimes u^{[r]}({\mathbf p}) + 
F_{h}^{[r-2]}({\mathbf r},t) \otimes \frac{p^{2}}{2m} 
u^{[r-2]} ,
\end{eqnarray}
where
\begin{eqnarray}
u^{[r]} = [ {\mathbf u} \hdots r-times \hdots {\mathbf u} ] ,
\end{eqnarray}
is the $r$-rank tensor resulting from the inner tensorial product of
$r$ characteristics velocities
\begin{eqnarray}
{\mathbf u} = {\mathbf p} / m .
\end{eqnarray}
Finally, defining the generalized fluxes operators 
\begin{eqnarray}
\hat{I}_{n}^{[r]}({\mathbf r}) = \int d^{3}p \, u^{[r]} \, 
\hat{n}_{1}({\mathbf r},{\mathbf p}) , 
\end{eqnarray}
\begin{eqnarray}
\hat{I}_{h}^{[r]}({\mathbf r}) = \int d^{3}p \, \frac{p^{2}}{2 \, m}
\, u^{[r]} \, \hat{n}_{1}({\mathbf r},{\mathbf p}) , 
\end{eqnarray}
and taking into account that
\begin{eqnarray}
\hat{n}_{2}({\mathbf r},{\mathbf p},{\mathbf r}',{\mathbf p}') = 
\hat{n}_{1}({\mathbf r},{\mathbf p}) \, 
\hat{n}_{1}({\mathbf r}',{\mathbf p}')  ,
\end{eqnarray}
we can write the auxiliary statistical operator of Eq.(6).
\section{Thermo-Hydrodynamic Variables for System with Arbitrary 
Energy-Dispersion Relation}
As shown elsewhere \cite{a26} at a quantum-mechanical description the
fluxes of energy (index $h$) and of matter (index $n$) are in
reciprocal space (of wavevectors ${\mathbf Q}$) given by 
\begin{eqnarray}
I_{h}^{[r]}({\mathbf Q},t) = \sum_{{\mathbf k}} \frac{1}{2} 
\left( \varepsilon_{{\mathbf k}+\frac{1}{2}{\mathbf Q}} + 
\varepsilon_{{\mathbf k}-\frac{1}{2}{\mathbf Q}} \right) \, 
u^{[r]}({\mathbf k},{\mathbf Q}) \, n_{{\mathbf k}{\mathbf Q}}(t) , 
\end{eqnarray}
\begin{eqnarray}
I_{n}^{[r]}({\mathbf Q},t) = \sum_{{\mathbf k}}  
u^{[r]}({\mathbf k},{\mathbf Q}) \, n_{{\mathbf k}{\mathbf Q}}(t) , 
\end{eqnarray}
where 
\begin{eqnarray}  
u^{[r]}({\mathbf k},{\mathbf Q}) = 
[ {\mathbf u}({\mathbf k},{\mathbf Q}) \hdots r-times \hdots 
{\mathbf u}({\mathbf k},{\mathbf Q}) ] , 
\end{eqnarray}
is a rank-$r$ tensor composed by the tensorial product of $r$-times
the generating velocity
\begin{eqnarray}  
{\mathbf u}({\mathbf k},{\mathbf Q}) = 
\frac{1}{\hbar} \nabla_{{\mathbf k}} \varepsilon_{{\mathbf k}} + 
\sum_{l=1}^{\infty} \frac{1}{(2l+1)!} 
\left( \frac{1}{2} {\mathbf Q} \cdot \nabla_{{\mathbf k}} \right)^{2l}
\frac{1}{\hbar} \nabla_{{\mathbf k}} \varepsilon_{{\mathbf k}} .
\end{eqnarray}
We recall that $\varepsilon_{{\mathbf k}}$ is the energy dispersion
relation of the single particles, and 
\begin{eqnarray}  
n_{{\mathbf k}{\mathbf Q}}(t) = Tr \left\{ 
c_{{\mathbf k}+\frac{1}{2}{\mathbf Q}}^{\dagger} \, 
c_{{\mathbf k}-\frac{1}{2}{\mathbf Q}} \, 
\bar{\rho}(t,0) \right\} ,
\end{eqnarray}
is the average over the nonequilibrium ensemble of
Dirac-Landau-Wiegner single-particle dynamical operator. It can be
noticed in Eq.(D4) that the generating velocity is composed of a first
contribution consisting of the group velocity of the wavepacket of a
single particle in state $| {\mathbf k} \rangle$, and the following
terms - depending on ${\mathbf Q}$ - can be considered as
contributions indicating the change from point to point of the group
velocity, once in direct space we can write 
\begin{eqnarray}
{\mathbf u}({\mathbf k},{\mathbf r}) = \left\{  
\frac{1}{\hbar} \nabla_{{\mathbf k}} \varepsilon_{{\mathbf k}} + 
\frac{1}{24} \left( \nabla \cdot \nabla_{{\mathbf k}} \right)^{2}
\frac{1}{\hbar} \nabla_{{\mathbf k}} \varepsilon_{{\mathbf k}} + 
\hdots \right\} \, \delta( {\mathbf r} ) , 
\end{eqnarray}
where $\nabla$ is the gradient in ${\mathbf r}$-space, and we can see
that this expression contains a series of even powers of such
gradient. These expressions in a classical mechanical levels retain
the form except that we must introduce the linear momentum 
${\mathbf p}$ in place of $\hbar \, {\mathbf k}$. 
\section{The Kinetic Theory}
The kinetic theory that can be derived out of the nonequilibrium
ensemble formalism can be considered a fra-reaching generalization of
Mori's approach, involving nonlinearity, all order of particle
collisions, and covering arbitrarily fra-from-equilibrium
situations. The presentation unavoidably involves a lengthy and
somehow cumbersome mathematical handling, but we call the attention to
the simplest form (the Markovian approximation) which is usually the
one to be used, once in most problems are verified the restrictions
that, say, validate it \cite{a26,a42,a43}. 

We begin recalling the separation of the total Hamiltonian as given by
\begin{eqnarray}
\hat{H} = \hat{H}_{0} + \hat{H}^{\prime} ,
\end{eqnarray}
where $\hat{H}^{\prime} = \hat{H}_{1} + \hat{W}$, with $\hat{H}_{1}$
containing the internal interactions and $\hat{W}$ those with the
reservoir and eventual external sources of pertubation, while
$\hat{H}_{0}$ is the kinetic energy operator. Moreover, let us call in
a generic way as $\left\{ \hat{P}_{j} \right\}$, $j = 1, 2, \hdots$
the set of basic variables, which for the case we are considering on
the main text are those of Eq.(26), for which stands the selection
rule
\begin{eqnarray}
\left\{ \hat{P}_{j} , \hat{H}_{0} \right\} = 
\sum_{m} \alpha_{jm} \, \hat{P}_{m} , 
\end{eqnarray}
where $\alpha_{jm}$ are $c$-numbers or differential operators
depending on the represdentation being used, and in the case of the
variables of Eq.(26) we have that 
\begin{eqnarray}
\left\{ \hat{I}_{p}^{[r]}({\mathbf r}) , \hat{H}_{0} \right\} = 
-\nabla \cdot \hat{I}_{p}^{[r+1]}({\mathbf r}) ,  
\end{eqnarray}
with $p = h$ or $n$ and $r = 0, 1,2, \hdots$. We also recall that for
the nonequilibrium statistical operator of the system holds the
separation into two parts, as given by Eq.(22) namely
\begin{eqnarray}
\rho_{\epsilon}(t) = \bar{\rho}(t,0) + \rho_{\epsilon}^{\prime}(t) ,   
\end{eqnarray}
consisting of the ''instantaneously frozen'' $\bar{\rho}$, and the
part $\rho_{\epsilon}^{\prime}$ responsible for the irreversible
evolution of the macrostate of the system. 

The equations of evolution for the basic variables are the average
over the nonequilibrium ensemble of Hamilton equation of motion for
the basic dynamical variables, namely 
\begin{eqnarray}
\frac{d \hfill}{dt} Q_{j}(t) = \frac{d \hfill}{dt}Tr \left\{ 
\hat{P}_{j} \, \rho_{\epsilon}(t) \times \rho_{R} \right\} = 
Tr \left\{ \frac{1}{{\textrm i} \, \hbar} 
\left[ \hat{P}_{j} , \hat{H} \right] \, 
\rho_{\epsilon}(t) \times \rho_{R} \right\} ,
\end{eqnarray}
where $Tr$ is to be understood as integration over the composite phase
space of system and reservoir. Using Eqs.(D1) to (D4) we can
alternatively write that
\begin{eqnarray}
\frac{d \hfill}{dt} Q_{j}(t) &=&  
Tr \left\{ \frac{1}{{\textrm i} \, \hbar} 
\left[ \hat{P}_{j} , \hat{H}_{0} \right] \, 
\bar{\rho}(t,0) \times \rho_{R} \right\} + 
Tr \left\{ \frac{1}{{\textrm i} \, \hbar} 
\left[ \hat{P}_{j} , \hat{H}_{0} \right] \, 
\rho_{\epsilon}^{\prime}(t) \times \rho_{R} \right\} \nonumber \\
&&+ 
Tr \left\{ \frac{1}{{\textrm i} \, \hbar} 
\left[ \hat{P}_{j} , \hat{H}^{\prime} \right] \, 
\bar{\rho}(t,0) \times \rho_{R} \right\} + 
Tr \left\{ \frac{1}{{\textrm i} \, \hbar} 
\left[ \hat{P}_{j} , \hat{H}^{\prime} \right] \, 
\rho_{\epsilon}^{\prime}(t) \times \rho_{R} \right\} , 
\end{eqnarray}
But the second term on the right-hand side of this Eq.(D6) is null; in
fact 
\begin{eqnarray}
Tr \left\{ \frac{1}{{\textrm i} \, \hbar} 
\left[ \hat{P}_{j} , \hat{H}_{0} \right] \, 
\rho_{\epsilon}^{\prime}(t) \times \rho_{R} \right\} = 
\sum_{jm} \alpha_{jm} \, Tr \left\{ \hat{P}_{m} 
\rho_{\epsilon}^{\prime}(t) \times \rho_{R} \right\} = 0 , 
\end{eqnarray}
once we take into account that
\begin{eqnarray}
Tr \left\{ \hat{P}_{m} \, \rho_{\epsilon}(t) \times \rho_{R} \right\}
= Tr \left\{ \hat{P}_{m} \, \bar{\rho}(t,0) \times \rho_{R} \right\} ,
\end{eqnarray}
[cf. Eq.(23)]. The remaining term can be handled through an elaborate
algebra to obtain that the kinetic equations acquire the form of a
far-reaching generalization of Mori's equations, namely
\begin{eqnarray}
\frac{d \hfill}{dt} Q_{j}(t) = J_{j}^{(0)} + J_{j}^{(1)} +
\sum_{m=2}^{\infty} \Omega_{j}^{(m)}(t) ,
\end{eqnarray}
where $J^{(0)}$ is the first contribution on the rightof Eq.(D6), 
$J^{(1)}$ the third followed by a series of partial collision 
operators involving increasing orders in the interaction strengths
($m= 1, 2, \hdots$), and containing memory effects, space
correlations, and are highly nonlinear in the basic variables (the
collision operators on the right-hand side of Eq.(E9) are dependent on
the intensive nonequilibrium thermodynamic variables, $F$'s in Eq.(2),
which are dependent on the basic variables through the constitutive
equations, cf. Eq.(19) or Eqs.(43) and (45)). 

In the extremely complicated set of coupled equations (E9), we can
single out the simplified form consisting of the markovian
approximation. The Markovian limit of the kinetic theory is of
particular relevance as a result that, for a large class of problems,
the interactions involved in $\hat{H}^{\prime}$ are weak and the use
of lowest order in the equations of motion constitutes an excellent
approximation of good practical value. In chapter 6 in Ref.[26] are
described several examples of its application for which there follows
an excellent agreement between the calculation and the experimental
data. By means of a different approach, E. B. Davies \cite{a44} has
shown that in fact the Markovian approach can be validated in the weak
coupling (in the interaction) limit (retaining only the quadratic
contribution).

Making it explicit, the Markovian equations in the kinetic theory are 
\begin{eqnarray}
\frac{d \hfill}{dt} Q_{j}({\mathbf r},t) = 
J_{j}^{(0)}({\mathbf r},t) + J_{j}^{(1)}({\mathbf r},t) + 
J_{j}^{(2)}({\mathbf r},t) ,
\end{eqnarray}
where we have introduced explicitly the eventual dependence on
position ${\mathbf r}$. Th e three terms on the right-hand side of
Eq.(D10) are 
\begin{eqnarray}
J_{j}^{(0)}({\mathbf r},t) = Tr \left\{ \frac{1}{{\textrm i} \, \hbar} 
\left[ \hat{P}_{j}({\mathbf r}) , \hat{H}_{0} \right] \, 
\bar{\rho}(t,0) \times \rho_{R} \right\} , 
\end{eqnarray}
\begin{eqnarray}
J_{j}^{(1)}({\mathbf r},t) = Tr \left\{ \frac{1}{{\textrm i} \, \hbar} 
\left[ \hat{P}_{j}({\mathbf r}) , \hat{H}^{\prime} \right] \, 
\bar{\rho}(t,0) \times \rho_{R} \right\} , 
\end{eqnarray}
\begin{eqnarray}
_{I}J_{j}^{(2)}({\mathbf r},t) = 
\left( \frac{1}{{\textrm i} \, \hbar} \right)^{2} \int_{-\infty}^{t}
dt'\, e^{\epsilon (t'-t)} \, \times Tr \left\{ 
\left[ \hat{H}^{\prime}(t'-t)_{0} , \left[ \hat{H}^{\prime} , 
\hat{P}_{j}({\mathbf r}) \right] \right] 
\bar{\rho}(t,0) \times \rho_{R} \right\} , 
\end{eqnarray}
\begin{eqnarray}
_{II}J_{j}^{(2)}({\mathbf r},t) = 
\frac{1}{{\textrm i} \, \hbar} \sum_{k} \int_{-\infty}^{t}
dt'\, e^{\epsilon (t'-t)} \, \times Tr \left\{ 
\left[ \hat{H}^{\prime}(t'-t)_{0} , \hat{P}_{k}({\mathbf r}) \right] 
\bar{\rho}(t,0) \times \rho_{R} \right\} 
\frac{\delta J_{j}^{(1)}({\mathbf r},t)}{\delta Q_{k}({\mathbf r},t)} ,
\end{eqnarray}
and we recall that $J_{j}^{(0)}$ and $J_{j}^{(1)}$, which in Mori's 
terminology \cite{a45} are called precession terms, are related to 
the non-dissipative part of the motion, while dissipative effects 
are accounted for $J^{(2)}$. In many cases, as a result of 
particular symmetries in $\hat{H}^{\prime}$ and $\bar{\rho}$, 
there follows that $J_{j}^{(1)} = 0$. 

Finally, we notice that in Eqs.(35) to (40), we do have in the
right-hand side that the collision operator is 
\begin{eqnarray}
J_{p}^{[r]}({\mathbf r},t) = J_{p}^{[r](1)}({\mathbf r},t) + 
\sum_{m\geq2} \Omega_{p}^{[r](m)}({\mathbf r},t) ,
\end{eqnarray}
in the general case, or 
\begin{eqnarray}
J_{p}^{[r]}({\mathbf r},t) \simeq J_{p}^{[r](1)}({\mathbf r},t) + 
J_{p}^{[r](2)}({\mathbf r},t) ,
\end{eqnarray}
in the Markovian regime \cite{a26,a42,a43}.
%
%
%

%
\end{document}